%% file: main.tex
\input{src/head}

\begin{document}

\title{Neuromorphic Silicon Neuron Controller for Adaptive Deep Brain Stimulation in Parkinson's Disease
}

\author{
Md Abu Bakr Siddique$^1$, Jakub Or\l{}owski$^2$, Yan Zhang$^3$, and Hongyu An$^1$\\[4pt]
{\small $^1$Department of Electrical and Computer Engineering, Michigan Technological University, Houghton, Michigan, USA}\\
{\small $^2$School of Electrical and Electronic Engineering, University College Dublin, Dublin, Ireland}\\
{\small $^3$Department of Biological Sciences, Michigan Technological University, Houghton, Michigan, USA}\\[2pt]
{\small $^1$msiddiq5@mtu.edu, $^2$jakub.orlowski@ucd.ie, $^3$yzhang49@mtu.edu, $^1$hongyua@mtu.edu}
}

\maketitle


\begin{figure*}[!b]
\centering
\begin{minipage}{1.00\textwidth}
\hrule
\vspace{3pt}
\footnotesize
This preprint has been accepted for publication in the proceedings of the ACM International Conference on Neuromorphic Systems 2026 (ICONS 2026).
\end{minipage}
\end{figure*}

\input{src/abstract}

\input{src/intro}

\input{src/background}

\input{src/methodology}

\input{src/results}

\input{src/Conclusion}

\bibliography{refs}
\bibliographystyle{IEEEtran}

\end{document}

%% file: src/head.tex
\documentclass[conference]{IEEEtran}
\IEEEoverridecommandlockouts
\usepackage{cite}
\usepackage{amsmath,amssymb,amsfonts}
\usepackage{cuted}
\usepackage{algorithmic}
\usepackage{graphicx}
\usepackage{textcomp}
\usepackage{xcolor}
\usepackage{booktabs}
\usepackage{caption}
\usepackage{subcaption}
\usepackage{longtable}
\usepackage{array} 
\usepackage{xcolor}
\definecolor{darkgreen}{rgb}{0.0, 0.5, 0.0}
\usepackage{makecell}
\usepackage[table]{xcolor}
\usepackage{multirow}
\usepackage{graphicx}
\usepackage{siunitx}
\usepackage{float}
\usepackage{tabularx}
\usepackage{enumitem}
\usepackage{hyperref}
\usepackage{caption}
\usepackage{bm}
\usepackage{siunitx}
\usepackage{stfloats}
\def\BibTeX{{\rm B\kern-.05em{\sc i\kern-.025em b}\kern-.08em
    T\kern-.1667em\lower.7ex\hbox{E}\kern-.125emX}}

%% file: src/abstract.tex
\begin{abstract}
Parkinson's disease (PD) affects millions worldwide and causes severe motor symptoms. Adaptive deep brain stimulation (aDBS) delivers physiologically informed stimulation that can track fluctuations in PD motor symptoms, enabling more intelligent DBS control. However, most existing aDBS approaches are primarily algorithm- and software-driven, with limited efforts toward circuit realization, particularly low-power and implantable integrated circuits. This paper presents the Silicon Leaky Integrate-and-Fire Deep Brain Stimulation (SiLIF-DBS) controller, a neuromorphic silicon neuron stimulator implemented with metal-oxide-semiconductor (CMOS) technology. For system-level evaluation, a simplified computational model of the SiLIF-DBS controller is derived and embedded within a Parkinsonian cortico-basal ganglia framework for closed-loop validation. The system is driven by beta-band subthalamic nucleus local field potentials (STN-LFPs), with their average rectified value (Beta ARV) used as the control biomarker. Our SiLIF-DBS controller for aDBS suppresses pathological beta activity while consuming only 25\% of the power required by open-loop stimulation and achieving a suppression efficiency of $5.85\%$/$\mu$W. Overall, our SiLIF-DBS controller achieves strong beta suppression at substantially reduced power, delivering high suppression efficiency that demonstrates it is a viable foundation for low-power implantable aDBS.

\end{abstract}


%% file: src/intro.tex
\section{INTRODUCTION}
Parkinson’s disease (PD) is the second most common neurodegenerative disorder and affects more than 1\% of the population above the age of 60 years old around the world~\cite{maserejian2020estimation}. PD is a progressive neurodegenerative disorder whose motor symptoms are linked to abnormal rhythmic activity in the cortico-basal-ganglia-thalamic loop, particularly elevated beta-band (13-30 Hz) oscillations in the subthalamic nucleus (STN)~\cite{little2013adaptive,guidetti2021clinical,ren2026parkinson}. Conventional deep brain stimulation (DBS) remains clinically effective, but it delivers stimulation continuously despite fluctuations in the pathological state, increasing battery drain, pulse-generator duty cycle, power consumption, and stimulation-related side effects~\cite{guidetti2021clinical}. These limitations have motivated the development of adaptive DBS (aDBS), in which neural biomarkers are monitored in real time and the stimulation parameters are optimized based on the symptom-related neural activity~\cite{little2013adaptive,arlotti2018eight,schmidt2024home,oehrn2024chronic}. The pathological beta activity (13-30 Hz) is not merely elevated on average but is often organized into bursts whose duration and amplitude correlate with clinical impairment~\cite{tinkhauser2017beta}. Thus, the controllers of aDBS system can be designed using STN local field potentials (STN-LFP), smoothed beta envelopes, burst metrics, and beta average-rectified value (Beta\_ARV) features~\cite{fleming2020simulation,fleming2020self,schmidt2024home,bahadori2023efficient}. The state-of-the-art (SOTA) methods and algorithms include threshold-based, proportional and proportional-integral, fuzzy, reinforcement learning, and machine learning~\cite{gorzelic2013model,camara2015fuzzy,lu2019application,liu2020neural,tinkhauser2021controlling,quan2024multi}. These studies demonstrate that closed-loop DBS can reduce stimulation demand while preserving therapeutic efficacy. However, most SOTA approaches remain primarily algorithmic and impose substantial computational overhead. For example, fuzzy algorithms increase control complexity, while reinforcement learning and neural network approaches require extremely high computational resources, making them less suitable for the ultra-low-power constraints of implantable medical devices.

In contrast, neuromorphic circuits are a promising candidate for aDBS because silicon neurons implement integration, leakage, thresholding, reset, and refractoriness directly in device physics rather than as instruction-level software operations~\cite{indiveri2011neuromorphic,livi2009current}. For chronically implanted neurostimulators, such analog dynamics offer compact state representation, low-latency response, and lower energy consumption. Recent PD-related neuromorphic studies have already demonstrated energy-efficient symptom monitoring and LIF-based closed-loop modulation~\cite{siddique2023monitoring,biswas2025energy}. Yet a rigorous refractory-enabled silicon neuron aDBS controller co-designed with a matched computational surrogate and validated in a Parkinsonian beta-control loop is still missing.


This paper addresses the gap by presenting a neuromorphic silicon neuron stimulator for aDBS in PD based on a refractory-enabled SiLIF-DBS controller primitive. Our SiLIF-DBS controller realizes the controller itself as a refractory-enabled silicon neuron, combining compact operation, physical interpretability, and direct hardware--software co-design for implantable mixed-signal implementation. The main contributions of this work are as follows:
\begin{enumerate}[topsep=0pt, itemsep=0pt, parsep=0pt, partopsep=0pt]
    \item We design and tune a transistor-level SiLIF-DBS controller with explicit membrane, leak, threshold, reset, and refractory subcircuits.
    \item We derive a matched computational surrogate that preserves the same two-state membrane--refractory structure and fitted conductance relations, enabling rapid yet circuit-interpretable controller evaluation.
    \item We integrate the resulting controller into a Parkinsonian control loop driven by beta-band filtered STN-LFP and Beta\_ARV.
    \item We validate the proposed framework at multiple levels, including neuron-rate programming over \SI{50}-\SI{250}{Hz}, hardware--software waveform matching, biomarker-level closed-loop behavior, target-dependent suppression-energy tradeoffs, and benchmarking against SOTA aDBS controllers.
\end{enumerate}

%% file: src/background.tex
\section{BACKGROUND AND RELATED WORKS}
Adaptive deep brain stimulation (aDBS) for PD combines neural biomarker extraction, feedback control, and implantable circuit realization. Pathological beta-band (13--30 Hz) activity in STN-LFPs is a widely used biomarker of Parkinsonian motor dysfunction~\cite{kuhn2005relationship,little2013adaptive,guidetti2021clinical,orlowski2023iterative,evers2024off}. Controller design has expanded from threshold-based and model-based strategies to fuzzy, reinforcement-learning, neural-network, amplitude-modulated, and multiscale approaches that seek improved suppression--performance tradeoffs~\cite{gorzelic2013model,camara2015fuzzy,fleming2020simulation,fleming2020self,lu2019application,liu2020neural,bahadori2023efficient,quan2024multi}. Neuromorphic mixed-signal circuits are a promising substrate for low-power, event-driven aDBS~\cite{indiveri2011neuromorphic,livi2009current}, yet a gap remains between algorithmic controller design and silicon control primitives for PD neuromodulation~\cite{siddique2023monitoring,biswas2025energy}.

STN beta activity is widely used as a biomarker for assessing the PD motor symptom~\cite{eusebio2011deep,little2013adaptive,guidetti2021clinical}. Burst-based analyses further showed that pathological beta is often intermittent rather than stationary, motivating adaptive stimulation that responds to pathological episodes instead of continuous delivery~\cite{he2023beta}. This has motivated controller inputs based on filtered LFP amplitude, burst duration, average rectified beta value, and combined biomarkers~\cite{fleming2020simulation,tinkhauser2021controlling,schmidt2024home}.

Early work on PD controllers studied model-based rational feedback control and fuzzy inference systems~\cite{gorzelic2013model,camara2015fuzzy}. Fleming \emph{et al.} compared on--off, dual-threshold, proportional, and proportional--Integral (PI) schemes in a Parkinsonian computational model and demonstrated that PI-like policies can balance suppression and stimulation  efficiency~\cite{fleming2020simulation}. Later studies extended this direction through self-tuning controllers, reinforcement learning, neural-network stimulation, and multi-timescale strategies~\cite{fleming2020self,lu2019application,liu2020neural,bahadori2023efficient,tinkhauser2021controlling,quan2024multi}. However, most of these approaches are developed and evaluated primarily in simulation, with limited consideration of hardware implementation and the constraints of implantable, low-power closed-loop DBS systems.

Neuromorphic systems have shown that conductance-based and integrate-and-fire dynamics can be implemented efficiently using compact analog circuits with intrinsic event-driven behavior~\cite{indiveri2011neuromorphic,livi2009current}. This is attractive for implantable neuromodulation, where the controller should be local, low-power, and tightly coupled to the sensor--stimulator loop. Recent PD neuromorphic work has demonstrated low-energy symptom monitoring and LIF-based closed-loop modulation with lower power than conventional approaches~\cite{siddique2023monitoring,biswas2025energy}. However, no cohesive study has yet started from a transistor-level silicon neuron, derived a reduced computational twin with matched states and conductance laws, and validated the full controller inside a Parkinsonian aDBS loop using biomarker analysis and clinically relevant tradeoff metrics. This work fills that gap.

%% file: src/methodology.tex
\begin{figure*}[t]
  \centering  \includegraphics[width=0.83\textwidth]{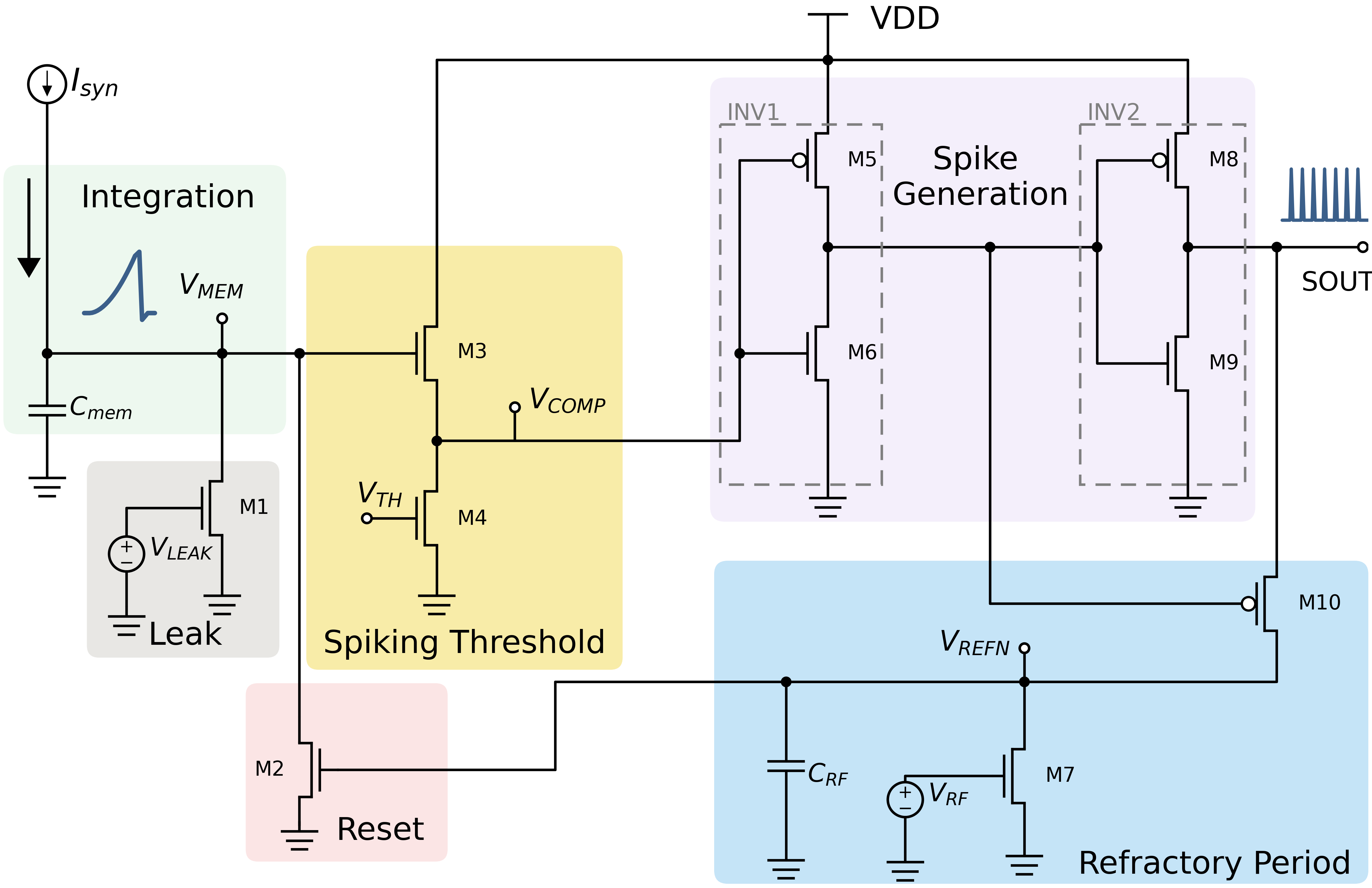}
  \caption{Proposed neuromorphic SiLIF-DBS controller. The input current $\bm{I}_{\mathrm{syn}}$ is integrated by $\bm{C}_{\mathrm{mem}}$ to generate $\bm{V}_{\mathrm{MEM}}$, with $\bm{M1}$ providing leak control through $\bm{V}_{\mathrm{LEAK}}$. The threshold stage $(\bm{M3},\bm{M4})$ compares $\bm{V}_{\mathrm{MEM}}$ with $\bm{V}_{\mathrm{TH}}$ to produce $\bm{V}_{\mathrm{COMP}}$, which is converted into the spike output $\bm{SOUT}$ by INV1 and INV2. The reset and refractory branch $(\bm{M2},\bm{M7},\bm{M10},\bm{C}_{\mathrm{RF}})$ generates $\bm{V}_{\mathrm{REFN}}$ using $\bm{V}_{\mathrm{RF}}$.}
  \label{fig:schematic_lif}
\end{figure*}

\section{Hardware--Software Co-Design of the Neuromorphic DBS Controller}
We adopt a hardware–software co-design methodology to develop the neuromorphic DBS controller. First, a silicon neuron controller is designed at the transistor level using CMOS technology, incorporating programmable membrane and refractory dynamics. Based on this neuromorphic controller, we then derive an equivalent computational model that preserves the same state variables, conductance mechanisms, threshold voltage, and spike-generation behavior. This model serves as an efficient surrogate for system-level simulation and validation and is further validated through direct comparison with the circuit in terms of membrane dynamics, refractory behavior, and spike timing. The purpose of this computational model is to enable its integration into Parkinsonian computational models \cite{little2013adaptive,fleming2020simulation,fleming2020self,biswas2025energy} to evaluate the effectiveness of the proposed neuromorphic controller in suppressing PD symptoms as well as its hardware performance.

\subsection{Neuromorphic SiLIF-DBS Controller Design}
 
 Our neuromorphic controller is built around a refractory-enabled silicon neuron with two capacitive state nodes: the membrane node $V_{\mathrm{MEM}}$ on capacitor $C_{mem}$ and the refractory node $V_{\mathrm{REFN}}$ on capacitor $C_{RF}$. As shown in Fig.~\ref{fig:schematic_lif}, the circuit contains five functional blocks: (1) synaptic-current integration at $V_{\mathrm{MEM}}$, (2) leak control through transistor M1 biased by $V_{LEAK}$, (3) threshold detection through the M3--M4 comparator stage producing $V_{\mathrm{COMP}}$, (4) spike generation through the cascaded inverters INV1 and INV2 producing the digital output \texttt{SOUT}, and (5) reset and refractory dynamics implemented through M2, M7, and M10. M2 is the reset transistor that discharges the membrane after spike generation, M7 is the refractory-discharge transistor controlled by $V_{RF}$, and M10 couples the spike-generation stage to the refractory branch. The refractory-control bias $V_{RF}$ is the main programming knob because it sets the effective refractory discharge strength, the refractory time constant, and hence the steady-state firing rate. 
 
 The proposed SiLIF-DBS controller is implemented using the SkyWater 130 nm PDK. The circuit includes membrane integration at $V_{\mathrm{MEM}}$ on $C_{mem}$, leak control through M1 biased by $V_{LEAK}$, threshold detection at $V_{\mathrm{COMP}}$, reset through M2, refractory storage at $V_{\mathrm{REFN}}$ on $C_{RF}$, refractory discharge through M7 biased by $V_{RF}$, and feedback coupling through M10. The spike output is \texttt{SOUT}. The circuit is used not merely as a neuron emulator but as the controller primitive itself. Within the PD loop, Beta\_ARV determines when the neuron generates DBS events, so the controller state is carried directly by the silicon neuron's membrane and refractory dynamics. Table~\ref{tab:transistor_sizes} lists the transistor sizes of the SiLIF-DBS controller.

We begin from the canonical current-based biological LIF neuron,
\begin{equation}
C_{mem}\frac{dV_{\mathrm{MEM}}}{dt}=I_{syn}(t)-g_L\bigl(V_{\mathrm{MEM}}-E_L\bigr),
\label{eq:bio_lif}
\end{equation}
and extend it with a time-varying refractory/adaptation conductance,
\begin{equation}
C_{mem}\frac{dV_{\mathrm{MEM}}}{dt}=I_{syn}(t)-\bigl(g_L+g_{ref}(t)\bigr)\bigl(V_{\mathrm{MEM}}-E_L\bigr),
\label{eq:bio_ref}
\end{equation}
whose decay between spikes is
\begin{equation}
\tau_{ref}\frac{dg_{ref}}{dt}=-g_{ref}.
\label{eq:gref_decay}
\end{equation}
$V_{\mathrm{MEM}}$ is the membrane voltage, $C_{mem}$ is the membrane capacitance, $I_{\mathrm{syn}}(t)$ is the synaptic input current, $g_L$ is the biological leak conductance, $E_L$ is the leak reversal potential, $g_{ref}(t)$ is the refractory or adaptation conductance, and $\tau_{ref}$ is its decay time constant. 

In the neuromorphic SiLIF-DBS controller, the membrane node is $V_{\mathrm{MEM}}$ on $C_{mem}$, driven by the current $I_{syn}(t)$. The effective membrane dynamics are written as

\begin{equation} C_{mem}\frac{dV_{\mathrm{MEM}}}{dt} = I_{\mathrm{syn}}(t)-\bigl(g_{leak}+g_{rst}(V_{\mathrm{REFN}})\bigr)V_{\mathrm{MEM}}, \label{eq:cmos_mem} 
\end{equation} 
where $g_{leak}$ is the effective leak conductance of M1 set by the bias $V_{LEAK}$, and $g_{rst}(V_{\mathrm{REFN}})$ is the effective reset or refractory shunt conductance seen by the membrane node through the M2--M10 reset/refractory path. We use the symbol $g_{rst}$, rather than a device-specific $g_{m2}$, because the effective shunt depends on the complete reset/refractory path and on the refractory state $V_{\mathrm{REFN}}$, not on transistor M2 alone. The comparator stage formed by M3--M4 produces the internal threshold node $V_{\mathrm{COMP}}$, and a spike is generated when the first inverter INV1 switches, i.e.,

\begin{equation}
V_{\mathrm{COMP}}\!\left(V_{\mathrm{MEM}},V_{TH}\right)\geq V_{\mathrm{trip,INV1}},
\label{eq:hw_spike}
\end{equation}
where $V_{\mathrm{trip,INV1}}$ is the effective switching threshold of INV1 and $V_{TH}$ is the threshold-bias input that helps determine $V_{\mathrm{COMP}}$. The refractory state is represented by the node $V_{\mathrm{REFN}}$ on capacitor $C_{RF}$. During and shortly after a spike, a charging path raises $V_{\mathrm{REFN}}$ toward the supply $V_{DD}$; between spikes, $V_{\mathrm{REFN}}$ decays through M7. Its dynamics are modeled as 

\begin{equation}
\small
C_{RF}\frac{dV_{\mathrm{REFN}}}{dt}
=
g_{chg}(t)\bigl(V_{DD}-V_{\mathrm{REFN}}\bigr)-g_{m7}(V_{RF})V_{\mathrm{REFN}},
\label{eq:cmos_ref}
\end{equation}
where $g_{chg}(t)$ is the effective spike-triggered charging conductance of the refractory branch, nonzero only during the refractory-charge phase, and $g_{m7}(V_{RF})$ is the effective M7 refractory-discharge conductance controlled by the refractory-bias voltage $V_{RF}$. The corresponding refractory time constant is 

\begin{equation} 
\tau_{ref}\approx \frac{C_{RF}}{g_{m7}(V_{RF})}. \label{eq:tau_ref} 
\end{equation}

\setlength{\textfloatsep}{1pt}
\begin{table}[!t]
\caption{Transistor sizes of the SiLIF-DBS Controller.}
\label{tab:transistor_sizes}
\centering
\small
\begin{tabular}{c@{\hspace{38pt}}c@{\hspace{38pt}}c}
\toprule
Device & Type & $W/L$ ($\mu$m/$\mu$m) \\
\midrule
$M1$  & nfet\_01v8 & $1.20 / 0.50$ \\
$M2$  & nfet\_01v8 & $1.20 / 0.50$ \\
$M3$  & nfet\_01v8 & $1.20 / 0.15$ \\
$M4$  & nfet\_01v8 & $1.20 / 0.15$ \\
$M5$  & pfet\_01v8 & $0.60 / 0.15$ \\
$M6$  & nfet\_01v8 & $0.60 / 0.15$ \\
$M7$  & nfet\_01v8 & $0.50 / 0.15$ \\
$M8$  & pfet\_01v8 & $1.20 / 0.25$ \\
$M9$  & nfet\_01v8 & $1.20 / 0.25$ \\
$M10$ & pfet\_01v8 & $0.55 / 0.15$ \\
\bottomrule
\end{tabular}
\end{table}

\subsection{Computational Model of the SiLIF-DBS Controller}
\label{computational_silif_controller}
To enable efficient system-level evaluation within PD computational frameworks, we derive a computational model of the SiLIF-DBS circuit that can be embedded into a PD model for assessment. The SiLIF-DBS computational model retains the same two state variables, namely the membrane voltage $V_{\mathrm{MEM}}$ and refractory voltage $V_{\mathrm{REFN}}$, along with the same leak, reset, threshold, and refractory mechanisms as in the circuit design. Between spike events, the dynamics are governed by the following equations:

\begin{equation} 
C_{mem}\frac{dV_{\mathrm{MEM}}}{dt} = I_{\mathrm{syn}}(t)-\bigl(g_{leak}+g_{rst}(V_{\mathrm{REFN}})\bigr)V_{\mathrm{MEM}}, 
\label{eq:comp_mem} 
\end{equation} 

\begin{equation} 
C_{RF}\frac{dV_{\mathrm{REFN}}}{dt} = -g_{m7}(V_{RF})V_{\mathrm{REFN}}, 
\label{eq:comp_ref_decay} 
\end{equation} 
with the event trigger 
\begin{equation} 
V_{\mathrm{MEM}}(t_k^-)\geq V_{TH}, 
\label{eq:comp_spike} 
\end{equation} 
where $t_k$ is the $k$th spike time. At each spike time $t_k$, the model resets $V_{\mathrm{MEM}}\leftarrow 0$, emits a spike event \texttt{SOUT}, and applies a refractory kick.

\begin{equation} 
V_{\mathrm{REFN}}\leftarrow \max\!\left(V_{\mathrm{REFN}},V_{ref,kick}\right), \label{eq:ref_kick} 
\end{equation} 
where $V_{ref,kick}$ is the spike-triggered refractory increment. The continuous-time surrogate is integrated with a fixed numerical step $\Delta t=\SI{0.5}{\micro s}$. To match the SiLIF-DBS circuit, the effective conductances are 

\begin{equation} 
g_{leak}=k_{leak}\bigl(V_{LEAK}-V_{TN,leak}\bigr)^2, \label{eq:gleak} 
\end{equation} 

\begin{equation} 
g_{rst}(V_{\mathrm{REFN}}) = g_{reset,max}\left(\frac{V_{\mathrm{REFN}}}{V_{DD}}\right)^p, \label{eq:grst} 
\end{equation} 
and
\begin{equation} 
\small
g_{m7}(V_{RF}) = k_{m7}\bigl(V_{RF}-V_{TN,m7}\bigr)^2+g_{floor}+g_{boost}(V_{RF}), 
\label{eq:gm7} 
\end{equation} 
where $V_{TN,leak}$ and $V_{TN,m7}$ are the effective threshold voltages of the leak and M7 fits, $k_{leak}$ and $k_{m7}$ are their scale factors, $g_{reset,max}$ is the maximum reset conductance, and $p$ is the nonlinearity exponent of the reset fit. The term $g_{floor}$ is a small positive floor conductance included to keep $g_{m7}(V_{RF})$ strictly positive and numerically stable at low $V_{RF}$. The additional low-bias correction term is 

\begin{equation} 
g_{boost}(V_{RF}) = g_{add}\, \mathrm{sat}\!\left( \frac{V_{knee}-V_{RF}}{V_{knee}-V_{low}};0,1 \right), 
\label{eq:gboost} 
\end{equation} 
where $g_{add}$ is the added low-bias conductance, $V_{knee}$ is the refractory-bias knee above which the correction is zero, $V_{low}$ is the lower fit breakpoint, and $\mathrm{sat}(x;0,1)=\min(\max(x,0),1)$.

\begin{figure}[t]
  \centering
 \includegraphics[width=0.48\textwidth]{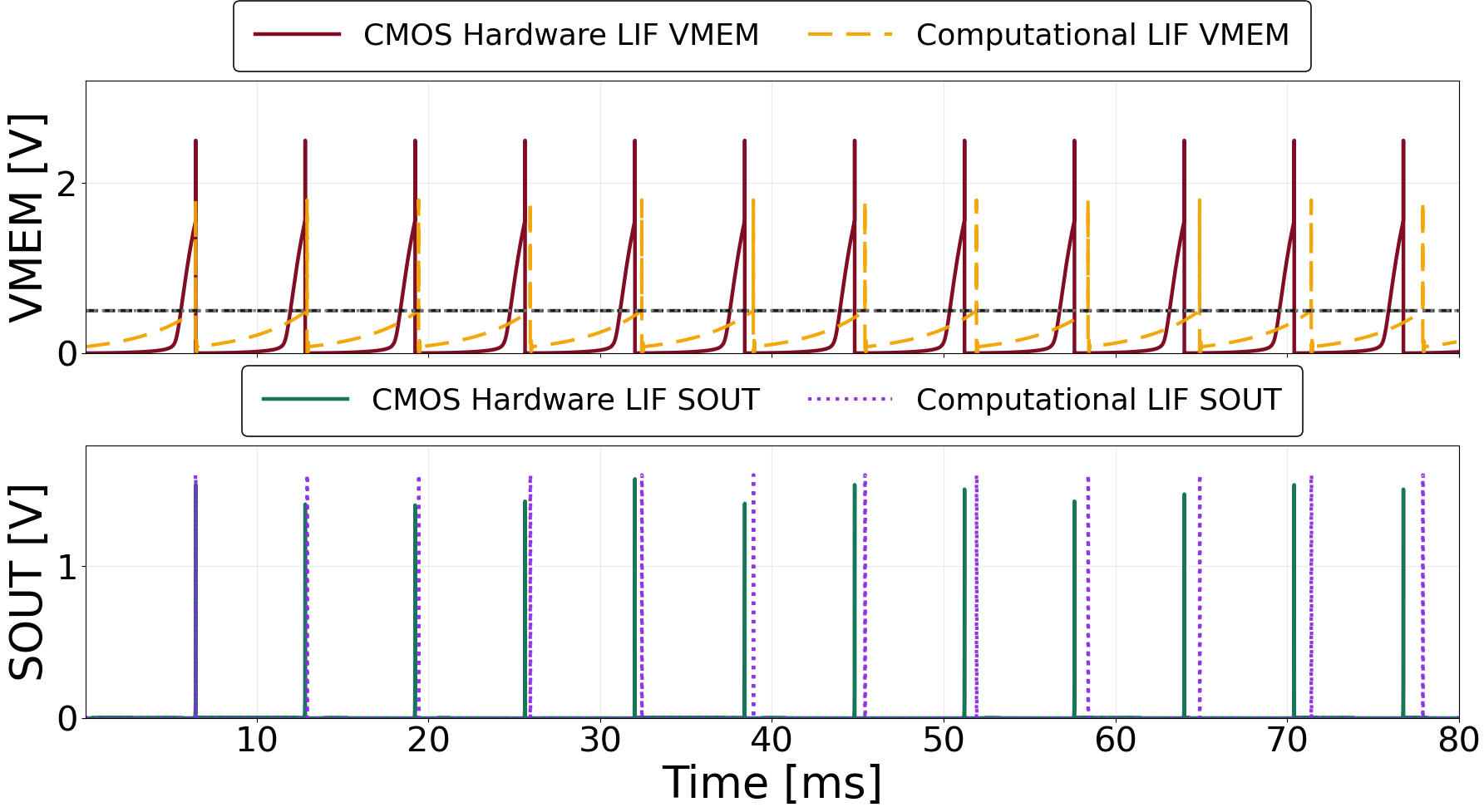}
  \caption{\boldmath Hardware--software equivalence of the SiLIF-DBS controller at \SI{150}{Hz} for $\bm{V}_{\bm{\mathrm{RF}}} = 0.26\ \mathrm{V}$.}
  \label{fig:match150}
\end{figure}

 Table~\ref{tab:fixedparams} summarizes the matching parameters for the circuit design and computational model of the SiLIF-DBS controller. Fig.~\ref{fig:match150} compares the neuromorphic SiLIF-DBS hardware controller and the matched computational surrogate at \SI{150}{Hz}. Under identical input current and matched parameters, the two show close agreement in membrane buildup, refractory decay, spike timing, and output spikes. This result demonstrates that the computational model of the SiLIF-DBS controller accurately captures the key firing behavior of the circuit implementation. As a result, it can serve as an efficient surrogate model and be embedded into PD computational frameworks for system-level evaluation. 
 
\begin{figure}[t]
\centering
\includegraphics[width=0.50\textwidth]{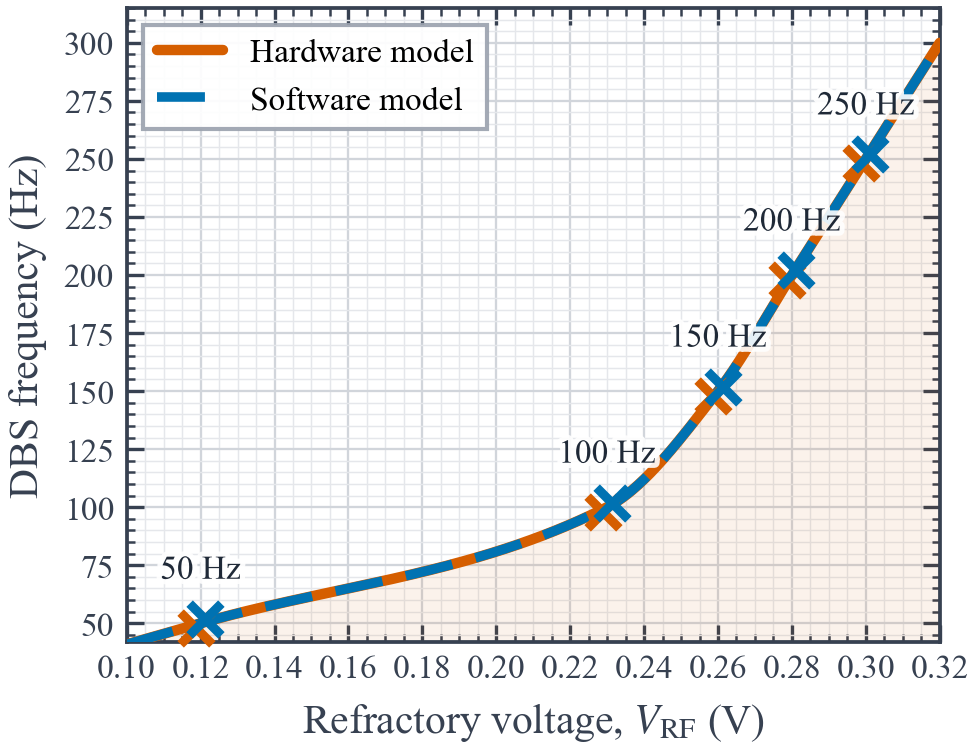}
\caption{Relationship between refractory voltage, $\bm{V}_{\bm{\mathrm{RF}}}$ and DBS frequency of the SiLIF-DBS controller.}
\label{fig:freq_tuning}
\end{figure}

 In addition, the stimulation frequency of our SiLIF-DBS can be tuned with the $V_{RF}$. Fig.~\ref{fig:freq_tuning} shows the neuron's firing rate increases as tuning voltage $V_{RF}$ becomes larger.


\begin{table}[t] 
\caption{Matching parameters for the circuit design and computational model of the SiLIF-DBS controller.} 
\label{tab:fixedparams} 
\centering 
\small 
\setlength{\tabcolsep}{3pt} 
\renewcommand{\arraystretch}{1.0} 
\begin{tabular}{p{1.5cm}p{2.4cm}p{4.5cm}} 
\toprule Parameter & Value & Role \\ 
\midrule 
$V_{DD}$ & \SI{1.8}{V} & Supply voltage \\ 
$V_{TH}$ & \SI{0.5}{V} & Comparator threshold bias \\ 
$V_{LEAK}$ & \SI{0.4}{V} & Leak bias for M1 \\ 
$C_{mem}$ & \SI{1.0}{pF} & Membrane capacitor \\ 
$C_{RF}$ & \SI{1.0}{pF} & Refractory capacitor \\ 
$I_{base}$ & \SI{10}{nA} & Base current for input scaling \\ 
$V_{ref,kick}$ & \SI{1.2}{V} & Refractory kick \\ 
$p_{SOUT}$ & \SI{20}{\micro s} & \texttt{SOUT} pulse width \\ 
$V_{TN,m7}$ & \SI{0.362}{V} & Effective threshold of M7 fit \\ $k_{m7}$ & $1.65\times 10^{-8}$ S/V$^2$ & Scale factor of M7 fit \\ 
$V_{TN,leak}$ & \SI{0.30}{V} & Effective threshold of leak fit \\ $k_{leak}$ & $4.0\times 10^{-9}$ S/V$^2$ & Scale factor of leak fit \\ 
$g_{reset,max}$ & $2.0\times 10^{-5}$ S & Maximum reset conductance \\ 
$p$ & $3.95$ & Exponent in $g_{rst}(V_{\mathrm{REFN}})$ \\ $g_{floor}$ & $3.0\times 10^{-15}$ S & Positive floor in $g_{m7}(V_{RF})$ \\ 
$g_{add}$ & $9.0\times 10^{-11}$ S & Added low-bias conductance\\ 
$V_{knee}$ & \SI{0.437}{V} & Knee of low-bias M7 correction \\ $V_{low}$ & \SI{0.20}{V} & Lower breakpoint of low-bias M7 \\ 
\bottomrule 
\end{tabular} 
\end{table}


\begin{figure}[t]
  \centering
 \includegraphics[width=0.48\textwidth]{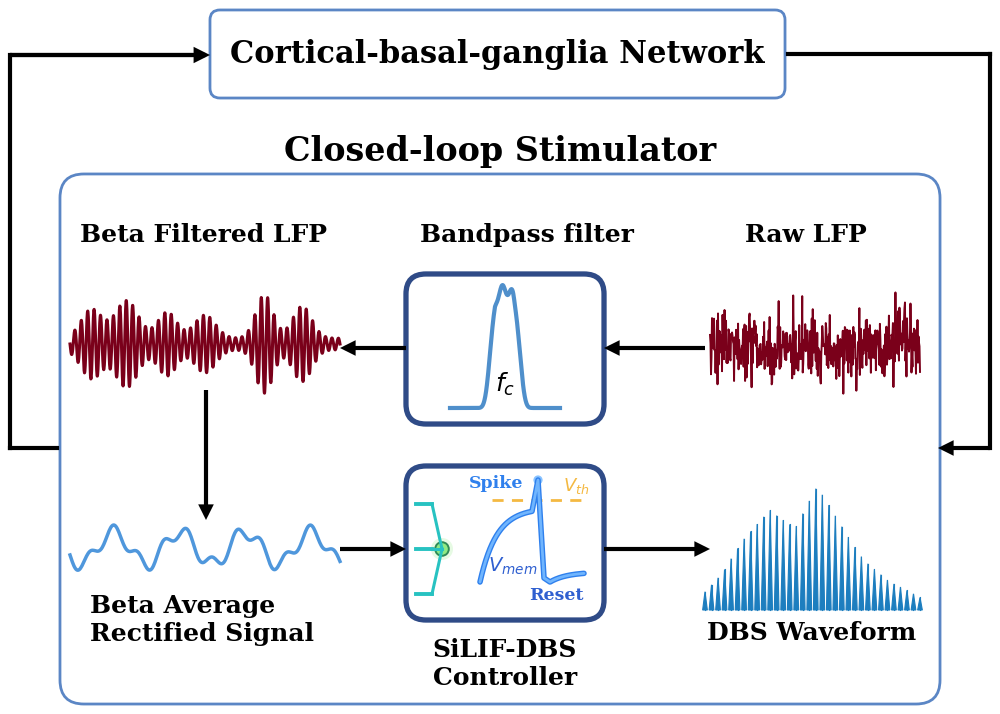}
  \caption{Illustration of integration of SiLIF-DBS controller into the PD computational model.}
  \label{fig:pipeline}
\end{figure}

\subsection{Embedding the SiLIF-DBS Controller into Parkinson's Disease Computational Model} 
\label{subsec:silif_embedding} 
In this section, we describe the integration of our SiLIF-DBS controller into a PD computational framework \cite{fleming2020simulation}.  Fig.~\ref{fig:pipeline} illustrates the architecture of the PD computational framework that consists of the critical cortico-basal-ganglia (CBG) network. The model follows a physiologically grounded structure, capturing the interaction between neural populations, pathological oscillations, and stimulation-induced modulation within the loop. In this framework, neural activity in the subthalamic nucleus (STN) is modeled through local field potentials (LFPs), which reflect population-level dynamics and serve as a measurable biomarker for PD pathological symptoms. The STN-LFP signal is processed through a beta-band filter (13–30 Hz) to extract pathological oscillatory components associated with Parkinson’s disease. The filtered signal is then rectified and averaged to compute the beta average rectified value (Beta\_ARV), which provides a representation of symptom-related neural activity. The Beta\_ARV is then fed into the pre-integrated PD controller within the computational model, such as a PID controller, to generate adaptive DBS stimulation signals. In this work, we replace the existing aDBS controller in the model with the computational model of the SiLIF-DBS controller developed in Section~\ref{computational_silif_controller}. 

During the $k$th controller interval, defined as the update window $[kT_s,(k+1)T_s)$ with update period $T_s=\SI{20}{ms}$, the continuous-time input in Eqs.~(\ref{eq:cmos_mem}) and~(\ref{eq:comp_mem}) is held constant, i.e., $I_{\mathrm{syn}}(t)=I_{\mathrm{syn},k}$ over that interval. The proportional scaling of this controller input uses the base current $I_{base}$ listed in Table~\ref{tab:fixedparams}. During each controller interval, the embedded SiLIF-DBS twin is advanced using Eqs.~(\ref{eq:comp_mem})--(\ref{eq:comp_spike}) together with the fitted conductance laws in Eqs.~(\ref{eq:gleak})--(\ref{eq:gboost}). Let $n_k$ denote the number of spikes generated during the $k$th controller interval $[kT_s,(k+1)T_s)$, and let $n_i$ denote the spike count in the $i$th interval. To stabilize the amplitude command, the controller forms a moving spike-count window of length $N_w=5$ controller samples, corresponding to a temporal window of $N_wT_s=\SI{100}{ms}$. The effective spike-rate estimate is 
\begin{equation} 
r_k = \alpha\,\frac{n_k}{T_s} + (1-\alpha)\,\frac{1}{N_wT_s}\sum_{i=k-N_w+1}^{k} n_i, 
\label{eq:spike_rate} 
\end{equation} 
where $r_k$ is the spike-rate estimate at the $k$th controller update, $n_k/T_s$ is the instantaneous spike rate in the current interval, $\frac{1}{N_wT_s}\sum_{i=k-N_w+1}^{k} n_i$ is the moving-window average rate, and $\alpha\in[0,1]$ controls the responsiveness--smoothing tradeoff; in this work, $\alpha=0.7$. The spike-rate estimate is mapped into the stimulation-amplitude command, 
\begin{equation} 
A_k=A_{\min}+\left(A_{\max}-A_{\min}\right)\, \mathrm{sat}\!\left(\frac{r_k}{r_{\max}};0,1\right), \label{eq:amp_map} 
\end{equation} 
where $A_k$ is the amplitude command applied during the $k$th controller interval, $A_{\min}=0$ mA and $A_{\max}=3.0$ mA are the allowable DBS amplitude bounds, and $r_{\max}=250$ Hz is the maximum calibrated operating rate used in the present controller. The saturation function $\mathrm{sat}(x;0,1)=\min(\max(x,0),1)$ prevents the normalized rate command from exceeding the available amplitude range. Finally, the DBS waveform delivered to the Parkinsonian model is generated by the same monophasic pulse generator used by the baseline controllers, ensuring a fair comparison at the waveform-delivery stage. Thus, SiLIF-DBS replaces the controller/amplitude-generation block, while the downstream pulse generator remains unchanged. The resulting stimulation signal is, 
\begin{equation} 
\small
u_{\mathrm{DBS}}(t)=A_k\sum_{m}\Pi\!\left(\frac{t-t_{k,m}}{p_{\mathrm{DBS}}}\right), \qquad t\in[kT_s,(k+1)T_s), \label{eq:dbs_waveform} 
\end{equation} 
where $u_{\mathrm{DBS}}(t)$ is the delivered DBS waveform, $\Pi(\cdot)$ denotes a unit rectangular pulse, $p_{\mathrm{DBS}}$ is the pulse width of the downstream DBS pulse generator, $\{t_{k,m}\}$ are the pulse times within the $k$th interval, and those pulse times are set by the fixed DBS carrier frequency $f_{\mathrm{DBS}}$.

\begin{figure}[t]
  \centering
  \includegraphics[width=0.47\textwidth]{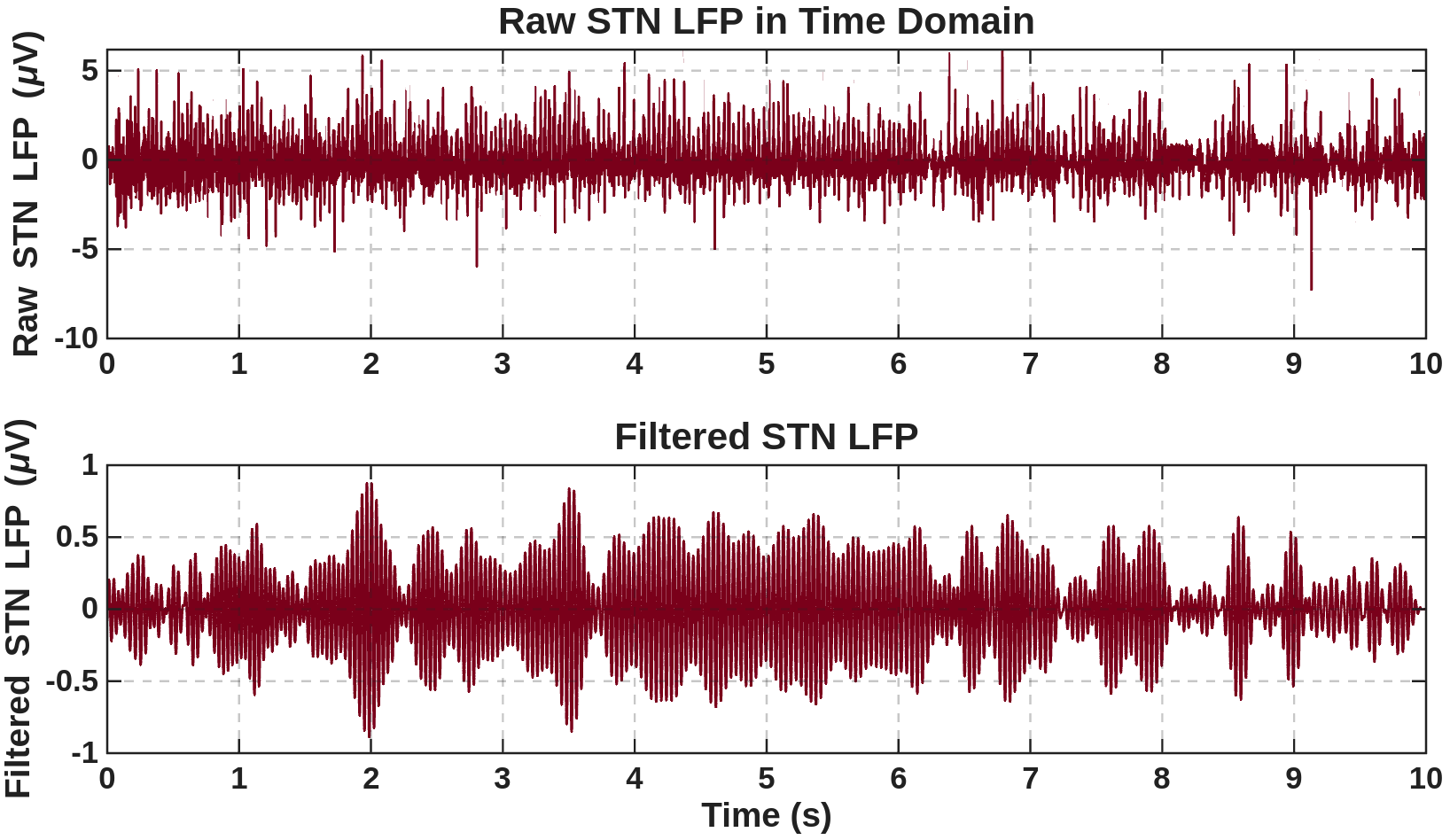}
  \caption{Raw STN-LFP and beta-band filtered STN-LFP for biomarker characterization.}
  \label{fig:biomarker_raw_filtered}
\end{figure}

\begin{figure}[t]
  \centering
  \includegraphics[width=0.48\textwidth]{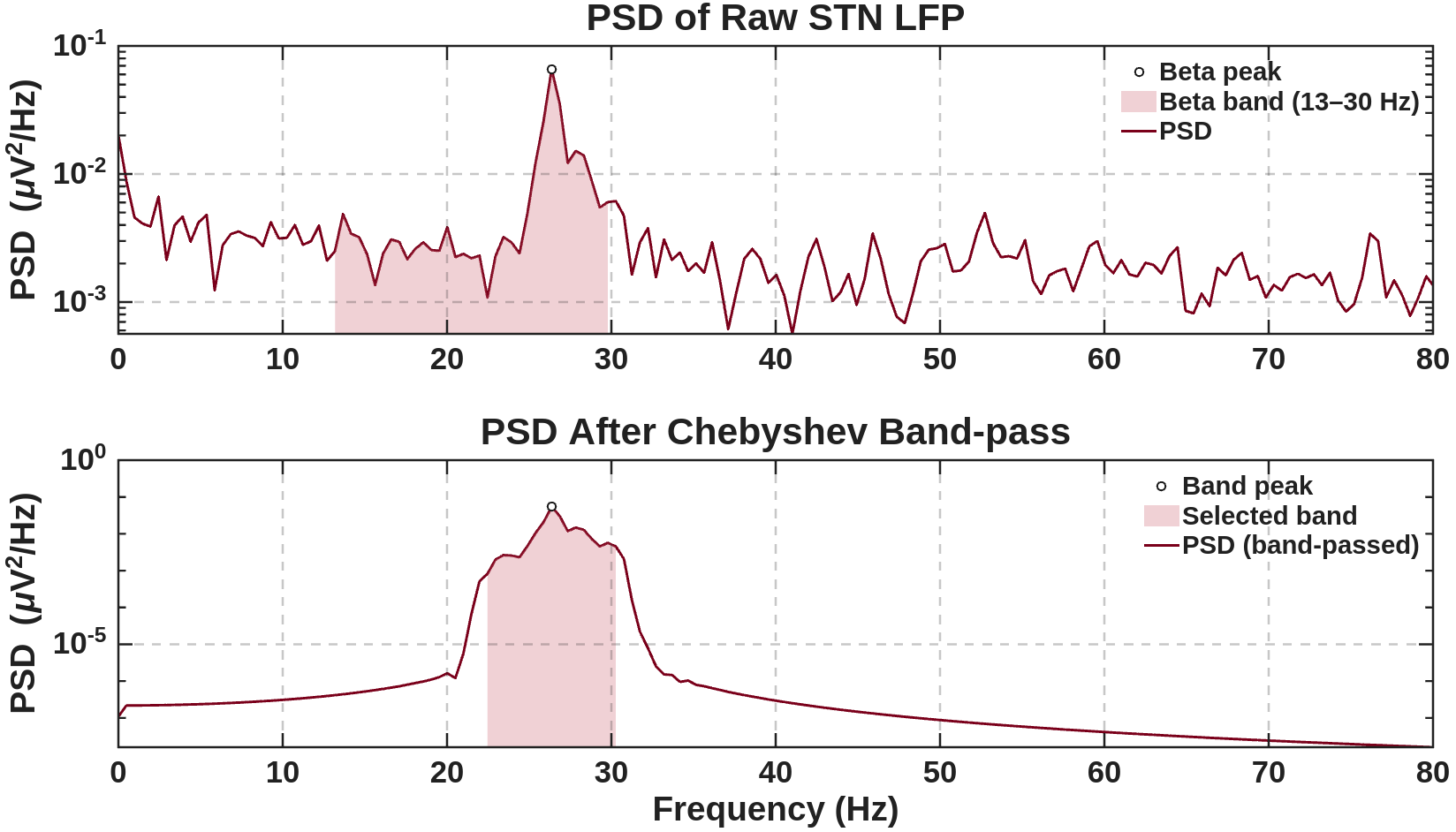}
  \caption{Power spectral density (PSD) of raw STN LFP and beta-band filtered STN-LFP.}
  \label{fig:biomarker_psd}
\end{figure}

The SiLIF-DBS controller is evaluated in the CBG framework under biomarker-driven closed-loop operation, where Beta\_ARV drives an amplitude-modulated DBS waveform reinjected into the model. Unlike continuous DBS, this strategy is event-driven and directly coupled to the pathological beta feature. Before evaluating the controller, we characterize the STN-LFP feedback biomarker. Fig.~\ref{fig:biomarker_raw_filtered}, and Fig.~\ref{fig:biomarker_psd} show that raw STN-LFP is broadband and not well suited for direct event-driven thresholding. After beta-band filtering, oscillatory packets become much clearer, and the corresponding power spectral density (PSD) exhibits a dominant peak in the selected beta-band (13--30 Hz) range. These observations justify using beta-band filtered STN-LFP and Beta\_ARV, rather than raw LFP, as the control signal.

%% file: src/results.tex
\section{Results and Discussion}

The SiLIF-DBS controller is evaluated by using a Parkinsonian cortico-basal-ganglia (CBG) computational model, where the controller is embedded as a closed-loop stimulation module. Specifically, the SiLIF-DBS controller receives beta-band (13–30 Hz) STN local field potentials (STN-LFPs) as input to generate the DBS stimulation signals. 

The performance of the proposed controller is evaluated using two assessment metrics: (1) power consumption, normalized to the open-loop DBS baseline, and (2) suppression efficiency, defined as the ratio between pathological beta suppression and energy usage. These metrics provide a direct measure of the tradeoff between control effectiveness and energy efficiency, which is critical for implantable aDBS systems. The average power consumption is calculated by the equation:
\begin{equation}
P_{\mathrm{avg}}=\frac{Z_E}{T_{\mathrm{sim}}}\int_{0}^{T_{\mathrm{sim}}} I_{\mathrm{DBS}}^{2}(t)\,dt,
\label{eq:power_avg}
\end{equation}
where $Z_E=0.5~\mathrm{k}\Omega$ is the electrode impedance, $I_{\mathrm{DBS}}(t)$ is stimulation current. 

The suppression efficiency is calculated by the equation: 
\begin{equation}
\small
\eta_{\mathrm{sup}}
=
\frac{100}{P_{\mathrm{avg}}}
\left(
1
-
\frac{1}{T_{\mathrm{sim}}}
\int_{0}^{T_{\mathrm{sim}}}
\frac{b_{\mathrm{DBS\ Off}}(t)-b_{\mathrm{ctrl}}(t)}
     {b_{\mathrm{DBS\ Off}}(t)}
\,dt
\right),
\label{eq:supp_eff}
\end{equation}
where $I_{\mathrm{DBS}}(t)$ is stimulation current, and $b_{\mathrm{DBS\ Off}}(t)$ and $b_{\mathrm{ctrl}}(t)$ are the DBS-off and controlled Beta\_ARV signals. 
The power consumption is normalized to open-loop DBS as a reference. Suppression efficiency is further quantified as the ratio of symptom suppression (\%) to power consumption ($\mu$W). This formulation enables a systematic comparison of both energy demand and suppression–energy trade-off across open-loop DBS, on–off, dual-threshold, and the proposed SiLIF-DBS controllers.

\begin{figure*}[t] 
\centering 
\includegraphics[width=1.00\textwidth]{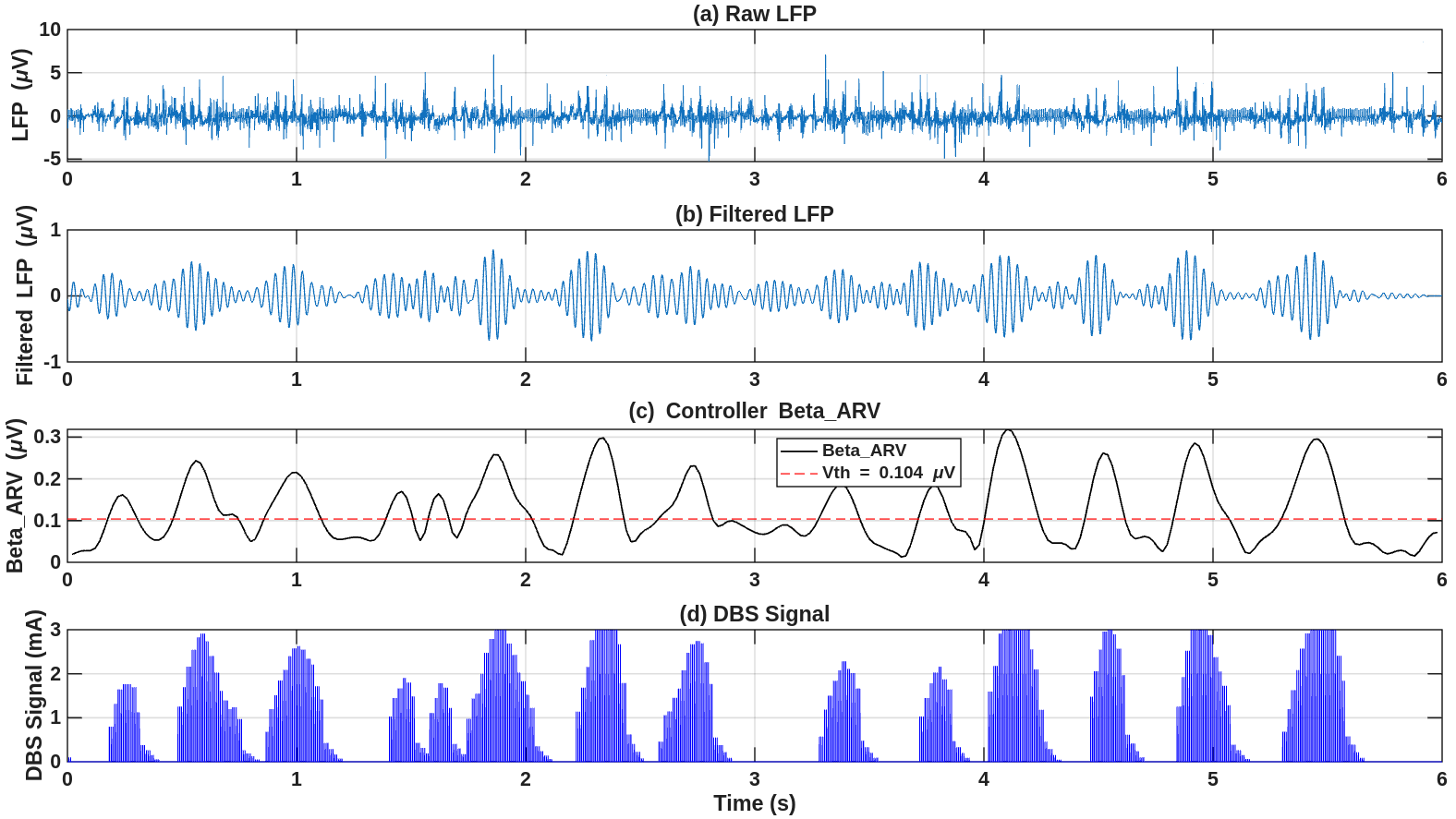} 
\caption{Output of the SiLIF-DBS controller with Beta\_ARV as input: (a) Raw STN-LFP, (b) Beta-band filtered STN-LFP, (c) Beta\_ARV, and (d) Adaptive DBS of the SiLIF-DBS controller.}

\label{fig:controller3s} 
\end{figure*}

Fig.~\ref{fig:controller3s} presents the output of the SiLIF-DBS controller that uses Beta\_ARV as the biomarker for input. Fig.~\ref{fig:controller3s}(a) shows the raw STN-LFP signal from the PD model, which is shown in Fig.~\ref{fig:pipeline}. The PD model generates the pathological neural activity of Parkinson's disease at beta band oscillation. Fig.~\ref{fig:controller3s}(b) illustrates the beta-band (13–30 Hz) filtered STN-LFP, highlighting the beta oscillation, which is the biomarker of Parkinsonian pathological symptoms. Fig.~\ref{fig:controller3s}(c) presents the corresponding Beta\_ARV, along with the predefined threshold for triggering the SiLIF-DBS controller. At last, Fig.~\ref{fig:controller3s}(d) shows the resulting DBS stimulation signal generated by the SiLIF-DBS controller. The stimulation is adaptively triggered in response to the Beta\_ARV dynamics, with pulses occurring only when the biomarker (Beta\_ARV) exceeds the threshold. This behavior arises from the controller design: the stimulation amplitude is governed by a spike-rate estimate in Eq.~(\ref{eq:spike_rate}) that combines the instantaneous spike activity with a short moving-window average. As a result, the spike history sustains the amplitude briefly after threshold crossing, leading to a smooth relaxation of stimulation instead of abrupt on–off switching. As a result, the magnitude of the DBS stimulation is modulated by the strength of the beta oscillations: stronger beta activity leads to higher stimulation amplitude, while weaker beta activity results in smaller stimulation amplitude. Fig.~\ref{fig:controller3s}(d) demonstrates that the controller can generate adaptive stimulation patterns, in contrast to conventional continuous stimulation with fixed amplitude.

\begin{table*}[!t]
\caption{Comparison of the state-of-the-art (SOTA) aDBS systems}
\label{tab:lit_summary}
\centering
\small
\setlength{\tabcolsep}{3pt}
\renewcommand{\arraystretch}{1.00}

\begin{tabular}{
>{\centering\arraybackslash}p{3.9cm}
@{\hspace{6pt}}
>{\centering\arraybackslash}p{3.3cm}
@{\hspace{6pt}}
>{\centering\arraybackslash}p{3.6cm}
@{\hspace{6pt}}
>{\centering\arraybackslash}p{4.2cm}
}
\toprule
\textbf{Representative aDBS works}
& \textbf{Control approach}
& \textbf{Implementation}
& \textbf{Neuromodulation parameters} \\
\midrule

\cite{gorzelic2013model,fleming2020simulation,quan2024multi}
& On-off / PI / PID
& Algorithms and simulation
& Amplitude, frequency \\

\cite{camara2015fuzzy}
& Fuzzy logic
& Algorithms and simulation
& DBS on/off decision \\

\cite{lu2019application,gao2020model}
& Reinforcement learning
& Algorithms and simulation
& Amplitude \\

\cite{liu2020neural,houston2019machine,oliveira2023machine}
& Machine learning
& Algorithms and simulation
& Amplitude \\

\textit{SiLIF-DBS} (Ours)
& Neuromorphic computing
& Analog CMOS circuit
& Frequency and amplitude \\

\bottomrule
\end{tabular}
\end{table*}

\begin{figure}[!t] 
\centering 
\includegraphics[width=0.50\textwidth]{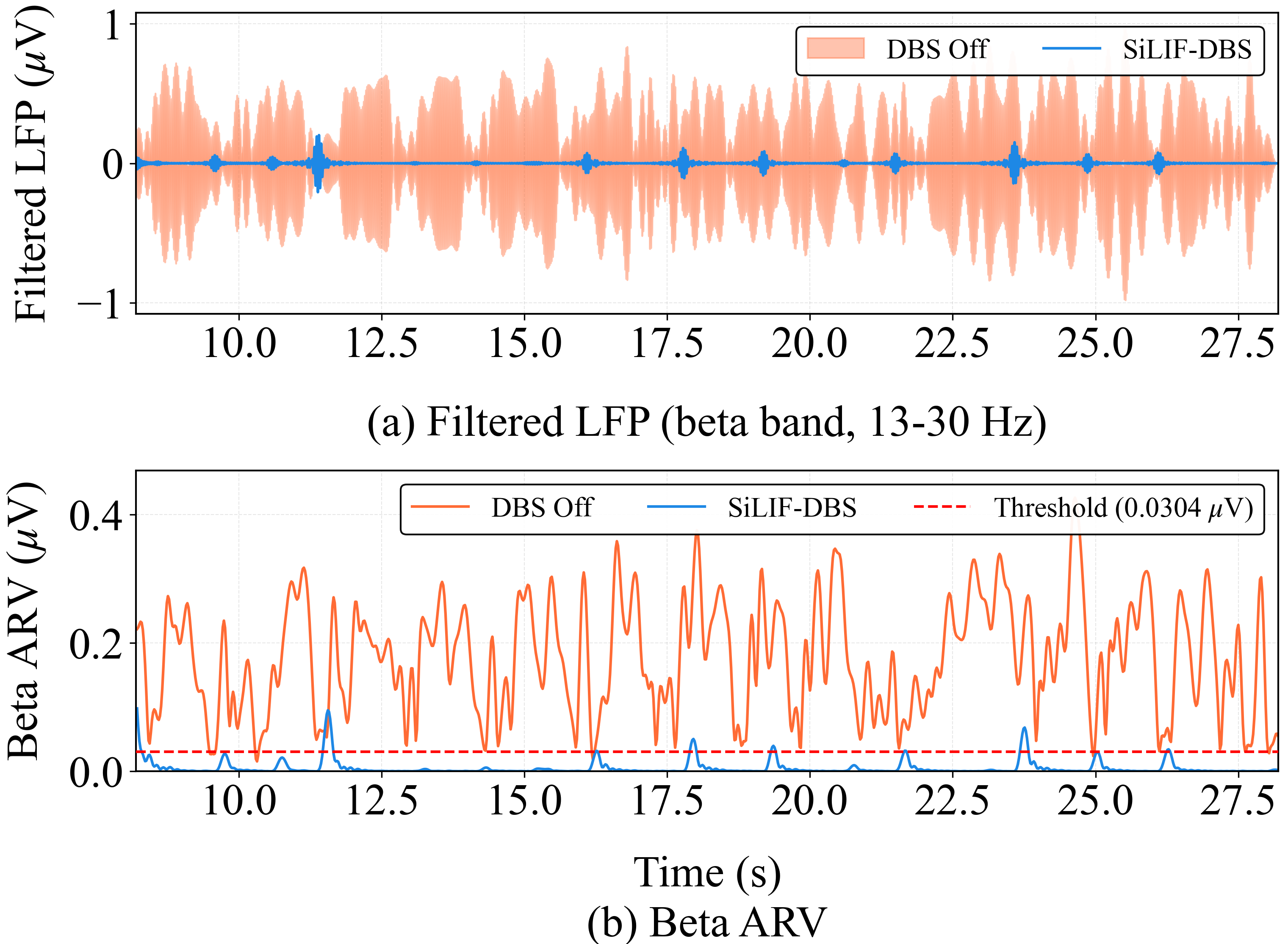}
\caption{Neuromorphic SiLIF-DBS suppresses pathological beta activity in the Parkinsonian model. 
(a) Beta-band (13–30 Hz) filtered STN local field potentials (LFPs) comparing DBS Off and SiLIF-DBS conditions.
(b) Beta average rectified value (Beta ARV) of the filtered LFP. }
\label{fig:validation} 
\end{figure}

Fig.~\ref{fig:validation} illustrates the suppression of pathological beta-band activity by the SiLIF-DBS controller in the Parkinsonian model. As shown in Fig.~\ref{fig:validation}(a), the beta-band (13–30 Hz) filtered STN-LFP exhibits strong oscillatory activity under the DBS-off condition, whereas these oscillations are significantly attenuated when SiLIF-DBS is applied. Correspondingly, Fig.~\ref{fig:validation}(b) shows the Beta\_ARV, where the DBS-off condition frequently exceeds the predefined threshold, indicating persistent pathological activity. In contrast, under SiLIF-DBS, the Beta\_ARV remains near or below the threshold.  In the DBS-off case, Beta\_ARV remains well above the threshold for much of the interval, while under SiLIF-DBS it is suppressed toward the threshold and only occasionally exceeds it. These results demonstrate that the SiLIF-DBS controller effectively suppresses pathological beta activity.

\begin{table}[t]
\caption{Power and efficiency comparison of aDBS controllers}
\label{tab:performance}
\centering
\small
\begin{tabular}{lcc}
\toprule
Controller & Power (\%) & Efficiency (\%/$\mu$W) \\
\midrule
Open-loop DBS (Baseline)& $100.0$ & $1.80$ \\
On--off & $41.0$ & $8.10$ \\
Dual-threshold & $52.0$ & $5.70$ \\
\textbf{SiLIF-DBS} (Ours) & $25.0$ & $5.85$ \\
\bottomrule
\end{tabular}
\end{table}

\begin{figure}[t]
  \centering
  \includegraphics[width=0.78\columnwidth]{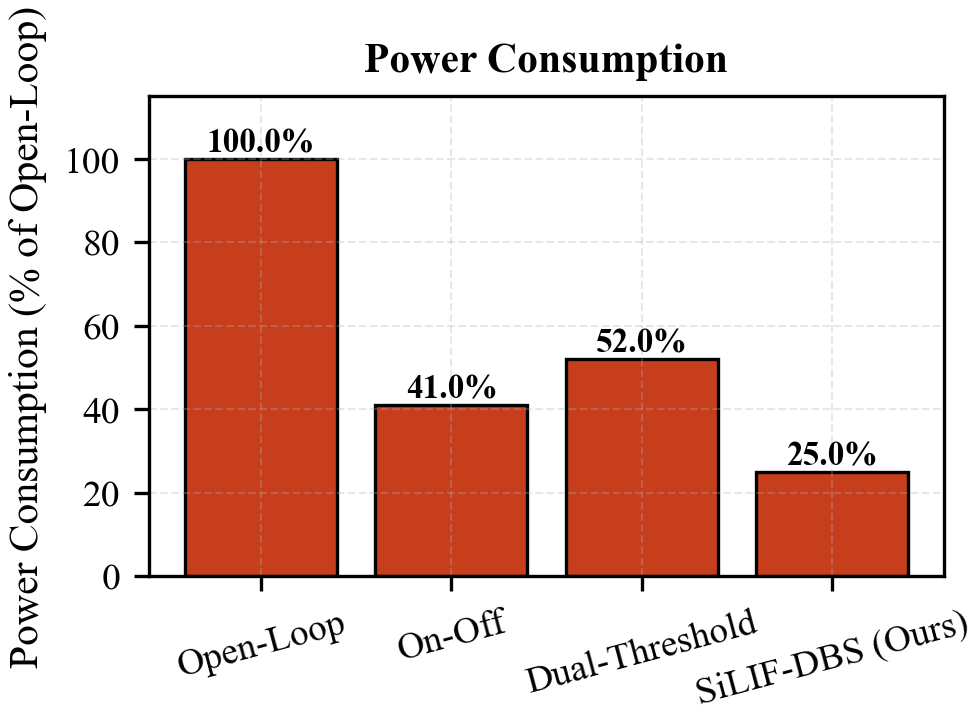}
  \caption{Energy Efficiency of SiLIF-DBS Compared to Conventional Controllers.}
  \label{fig:perf_power}
\end{figure}

\begin{figure}[t]
  \centering
  \includegraphics[width=0.78\columnwidth]{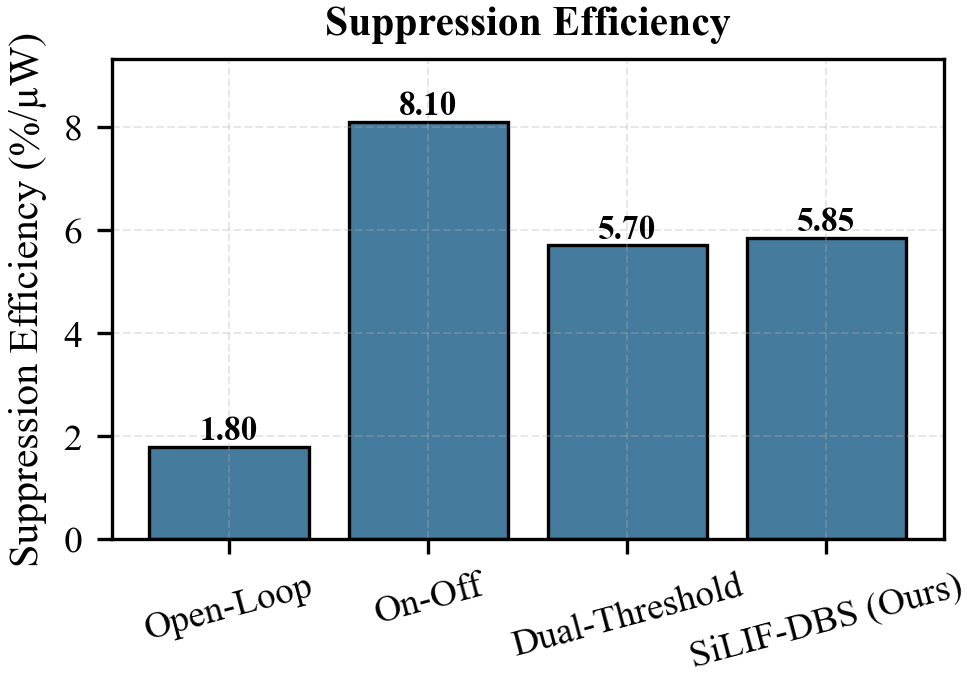}
  \caption{Suppression Efficiency of SiLIF-DBS and Conventional Controllers}
  \label{fig:perf_efficiency}
\end{figure}


Table~\ref{tab:performance}, Fig.~\ref{fig:perf_power}, and Fig.~\ref{fig:perf_efficiency} summarize the controller-level comparison. Open-loop DBS incurs the highest power cost because it stimulates aggressively and almost continuously. The on--off controller achieves the highest suppression efficiency, whereas the dual-threshold controller attains somewhat lower efficiency at higher power. SiLIF-DBS occupies an attractive intermediate regime, using only $25\%$ of open-loop power while achieving $5.85\%$/$\mu$W suppression efficiency, slightly above the dual-threshold controller and substantially above open-loop DBS. These results show that the proposed silicon-neuron controller achieves effective beta suppression at markedly lower power, making it a practical hardware-realizable alternative to conventional aDBS.

The results consistently support the proposed design of the neuromorphic SiLIF-DBS controller. Biomarker extraction isolates control-relevant beta activity; the transistor-level silicon neuron provides a controller tunable through a single refractory-control bias; and the matched computational surrogate preserves these dynamics for system-level exploration. In the Parkinsonian loop, the controller exhibits a clear suppression--energy tradeoff rather than an all-or-nothing response. This shows that the proposed approach not only produces closed-loop stimulation but also does so through a common physical mechanism that links neural state, controller dynamics, and stimulation demand. In particular, the same refractory process that shapes spike timing in the silicon neuron also determines how aggressively the controller responds to pathological beta activity, making the energy--suppression tradeoff an intrinsic property of the controller rather than a separate algorithmic add-on. For an implantable device, the most useful controller is therefore not necessarily the one with the smallest isolated error, but the one that preserves clinically meaningful suppression under battery, area, and thermal constraints. Thus, the refractory-enabled SiLIF-DBS controller is compelling because its internal state variables are directly realizable in analog hardware, and the same circuit parameters that govern spike generation also govern the energy--suppression tradeoff. This gives the controller a level of physical interpretability that is often absent in software-first aDBS strategies. The practical significance is that controller tuning can be understood not only at the behavioral level but also at the circuit level through membrane, leak, reset, and refractory mechanisms, which is especially valuable for future mixed-signal implant design. Table~\ref{tab:lit_summary} highlights this distinction by showing that SiLIF-DBS combines transistor-level realizability, hardware--software co-design, and closed-loop validation in a way that is not jointly emphasized in prior clinical, algorithmic, or neuromorphic aDBS approaches.


%% file: src/conclusion.tex
\section{CONCLUSION}
This paper presented SiLIF-DBS, a neuromorphic silicon neuron stimulator for adaptive DBS in Parkinson's disease based on a refractory-enabled SiLIF-DBS controller and a matched computational surrogate. The work unified three representations of the same controller: a biologically interpretable LIF model, a transistor-level silicon neuron with explicit refractory control, and a reduced computational surrogate preserving the same membrane and refractory states. Across \SI{50}-\SI{250}{Hz}, the hardware and computational models showed close waveform-level agreement. In Parkinsonian closed-loop simulations, the controller used beta-band filtered STN-LFP and Beta\_ARV to generate temporally sparse DBS bursts that suppressed pathological beta activity while substantially reducing power consumption relative to open-loop stimulation. The resulting suppression--energy tradeoff, together with the controller’s direct circuit realizability, supports refractory-enabled silicon neuron controllers as strong candidates for future low-power implantable aDBS systems.

\vspace{-2mm}
\section*{ACKNOWLEDGMENT} 
This work was supported by the program: Disability and
Rehabilitation Engineering (DARE) of the National Science Foundation under Award Number 2301589, and by Taighde Éireann--Research Ireland under Grant number 22/FFP-A/10954.